\documentclass[usenatbib]{mn2e}
\usepackage{graphicx}

\newcommand{\lesssim}{\mathrel{\hbox{\rlap{\hbox{\lower4pt\hbox{$\sim$}}}\hbox{$<$}}}}
\newcommand{\gtrsim}{\mathrel{\hbox{\rlap{\hbox{\lower4pt\hbox{$\sim$}}}\hbox{$>$}}}}
\newcommand{\beq}{\begin{equation}}
\newcommand{\eeq}{\end{equation}}
\newcommand{\beqa}{\begin{eqnarray}}
\newcommand{\eeqa}{\end{eqnarray}}

\newcommand{\ep}{E_{\rm p}}
\newcommand{\ee}{E_{\rm e}}
\newcommand{\rcut}{R_{\rm cut}}

\title[Hard X-ray emission of the Earth's atmosphere]{Hard X-ray
emission of the Earth's atmosphere: Monte Carlo simulations} 

\author[S. Sazonov et al.]{S. Sazonov$^{1,2}$\thanks{E-mail:
sazonov@mpa-garching.mpg.de}, E. Churazov$^{1,2}$, R. Sunyaev$^{1,2}$
and M. Revnivtsev$^{1,2}$\\ 
$^{1}$Max-Planck-Institut f\"ur Astrophysik,
Karl-Schwarzschild-Str. 1, 85740 Garching bei M\"unchen, Germany\\
$^{2}$Space Research Institute, Russian Academy of Sciences,
Profsoyuznaya 84/32, 117997 Moscow, Russia}

\begin{document}

\maketitle

\begin{abstract}
We perform Monte Carlo simulations of cosmic ray-induced hard X-ray
radiation from the Earth's atmosphere. We find that the shape of the
spectrum emergent from the atmosphere in the energy range 25--300~keV 
is mainly determined by Compton scatterings and photoabsorption, and
is almost insensitive to the incident cosmic-ray spectrum. We
provide a fitting formula for the hard X-ray surface brightness
of the atmosphere as would be measured by a satellite-born
instrument, as a function of energy, solar modulation level,
geomagnetic cutoff rigidity and zenith angle. A recent measurement by
the INTEGRAL observatory of the atmospheric hard X-ray flux during the
occultation of the cosmic X-ray background by the Earth  agrees with
our prediction within 10\%. This suggests that Earth observations
could be used for in-orbit calibration of future hard X-ray
telescopes. We also demonstrate that the hard X-ray spectra generated 
by cosmic rays in the crusts of the Moon, Mars and Mercury should be
significantly different from that emitted by the Earth's atmosphere.
\end{abstract}

\begin{keywords}
Earth -- gamma-rays: observations -- X-rays: diffuse
background -- Moon 
\end{keywords}

\section{Introduction}
\label{s:intro}

Starting in the 1960's, observations from balloons 
\citep{petetal73,danste74,schetal80} and spacecraft  
\citep{goletal74,schpet74,imhetal76,wilmah92,shamur01,shaetal03} have
shown that the Earth's atmosphere is bright in hard X-rays and
gamma-rays owing to its bombardment by cosmic 
rays of Galactic origin. This is the case despite the fact that the
energy deposited by the cosmic rays into the atmosphere is some nine orders of
magnitude smaller than the total energy budget coming to us from the
Sun and that only $\sim 1$\% of the energy deposited by the
cosmic rays is re-radiated into space as gamma-rays and much less as
hard X-rays.

Atmospheric hard X-ray emission is of interest not only because it
provides insight into the interaction of energetic
cosmic particles with air, but also in connection with the possibility
of measuring the spectrum of the cosmic X-ray background (CXB) using
the Earth as a shield. In January--February 2006 a series of
observations following this approach were performed by the INTEGRAL
orbital observatory \citep{chuetal07} and drew our attention to the
problem of atmospheric X-ray radiation. In observations of this kind,
cosmic ray-induced emission on the one hand causes significant
contamination of the CXB decrement (due to occultation by the Earth)
at energies $\sim 30$--100~keV, but on the other, could in
principle be used for calibrating a detector's response in the high-energy
region (substantially above 100~keV) where atmospheric emission
dominates over the CXB signal. During the INTEGRAL observations the
Earth covered $\sim 100$~sq. deg of the sky and was as bright as
$\sim 200$~mCrab in the spectral band 100--200~keV \citep{chuetal07}.

Despite the long history of experimental (see references above) and
theoretical \citep{puskin70,danste74,ling75,grasch77,deaetal89}
studying, there remains significant uncertainty regarding the spectrum
of atmospheric X-ray and gamma-ray radiation. However,
the nuclear and electromagnetic processes responsible for 
the generation and propagation of X- and gamma-rays in the atmosphere,
although complex, are well understood and already incorporated in
computer codes. Also the incident energy spectrum of Galactic
cosmic rays has been measured with high precision
(see e.g. \citealt{gaietal01} for a review) and there is a good
understanding of how it is modified below $\sim 10$~GeV by the magnetic
fields of the Solar system and Earth. All this suggests that the energy and
angular distributions of atmospheric X-ray and gamma-ray emission
should be predictable by simulations with fairly good ($\sim
10$\%) accuracy.

We have performed such a numerical modelling using the popular
toolkit {\sl GEANT} \citep{geant03}, focusing on making predictions
for observations of the Earth from space in the hard X-ray band
25--300~keV. We show below that the hard X-ray emission of the
Earth is characterized by a well-defined spectral shape
(Sect.~\ref{s:simul}) and that its intensity depends in a predictable way
on the phase of the solar cycle and on the geomagnetic rigidity map 
corresponding to a given observation (Sect.~\ref{s:predict}). 
Therefore the Earth could become a good absolute calibrator for future
orbital hard X-ray telescopes, in particular when used in combination with
accurate monitoring of the cosmic-ray spectrum by large-area,
high-resolution particle physics detectors such as the Alpha
Magnetic Spectrometer to operate on the International Space Station
\citep{baretal04}.

In addition we demonstrate (Sect.~\ref{s:planets}) that
the cosmic ray-irradiated solid crusts of the Moon, Mars and Mercury
should generate hard X-ray spectra different from the one produced in
the Earth's atmosphere. 

\section{Simulations}
\label{s:simul}

Our Monte Carlo code is based on the {\sl GEANT4} software package
(release 8.1), which was developed at CERN as an extension of {\sl
GEANT3} and has many applications across particle physics and astronomy
\citep{geant03}. We adopted the physics model
originally constructed for evaluating the effects of exposing the test
masses of the Laser Interferometer Space Antenna to energetic
particles in space \citep{araetal05}. This model includes a
state-of-the-art description (in particular, intra-nuclear cascade
models) of hadronic interactions of cosmic-ray protons and
$\alpha$-particles, and their air-shower products over many orders of
magnitude in energy, as well as the low-energy electromagnetic
package in {\sl GEANT4}, which extends the standard electromagnetic
package by carefully treating such processes as Rayleigh scattering,
photoabsorption and fluorescence, important in the
X-ray regime.

Unless otherwise stated, we use a plane-parallel approximation
for the Earth's atmosphere. Specifically, our adopted geometry is a
slab of height 96~km and of very large ($2\times 10^4$~km) width and
length. The air consists of atoms of nitrogen, oxygen
and argon with mass fractions of 75.5\%, 23.1\% and 1.3\%,
respectively. The air's density decreases exponentially with altitude from a
sea-level value of 1.2~mg~cm$^{-3}$ with a scale height of 8~km. The
total column density of our model atmosphere is thus $\sim
960$~g~cm$^{-2}$. This model is designed to approximate the
real structure of the Earth's atmosphere\footnote{See
e.g. http://modelweb.gsfc.nasa.gov} below $\sim 100$~km. The remaining
Earth's atmosphere above this level is so rarefied ($\sim
1$~mg~cm$^{-2}$) that only a tiny fraction ($\lesssim 10^{-4}$) of the
total X-ray and gamma-ray radiation escaping from the atmosphere into
space is produced there, since the bulk of the emergent
flux is produced in the upper several tens of g~cm$^{-2}$ of the
atmosphere (above $\sim 20$~km from the sea level).  

Importantly, by performing many test runs, we found that the
properties of the emergent hard X-ray radiation remain unchanged
within the statistical uncertainties if we 
increase or decrease severalfold the total height, scale height, or
the sea-level density of the atmosphere, as long as the total column 
density of the atmosphere remains higher than $\sim
200$~g~cm$^{-2}$. It is the chemical composition of the air that
largely determines the outcome of simulations. Furthemore, we 
also carried out a series of simulations using a realistic spherical
geometry for the atmosphere and convinced outselves that the plane-parallel
approximation is adequate for treating the problem under
consideration (see Sect.~\ref{s:simul_proton} below).

We emphasize that we use our simulations to formulate predictions
specifically for observations of the Earth from satellite orbits. This
implies that the presented results should not be applied to
observations performed from balloon (i.e. intra-atmospheric) altitudes.  

\subsection{Spectra induced by monoenergetic protons}
\label{s:simul_proton}

We first performed a series of simulations in which the atmosphere was
bombarded by monoenergetic protons. The protons were assumed to
come from a uniform background subtending the 2$\pi$ space above the
atmosphere. This implies that the flux density of incoming particles 
$dF/d\mu\propto \mu$, where $\mu=\cos\theta$ and $\theta$ is the
zenith angle.  We accumulated the energy spectra (above an
imposed threshold of 1~keV) and angular distributions of X- and
gamma-rays produced in the proton-induced cascades and emergent from the
atmosphere. Similar information was collected for the emergent electrons
and positrons, to later assess the role of re-entrant albedo particles.

We typically sampled (50,000--125,000)$/\ep$ protons in a single
run, where $\ep$ (GeV) is the proton kinetic energy. By performing a
number of test runs, we convinced ourselves that this number of sampled
particles enables recovering outgoing hard X-ray to gamma-ray spectra
(between $\sim 25$~keV and $\sim 10$~MeV) with a bin-to-bin statistical
noise of less than 5\% when a logarithmic grid of 10 energy
bins per dex is used, and with a scatter in the amplitudes of spectra
obtained in different runs of less than 3\%. These levels of statistical
uncertainty characterize all spectral plots presented below. We note
that the accuracy of our simulations becomes significantly worse at
photon energies below $\sim 25$~keV and above $\sim 10$~MeV (see in
particular Fig.~\ref{proton_setups} below), in the 
regions lying outside the focus of this paper. 

\begin{figure}
\centering
\includegraphics[height=0.5\textheight]{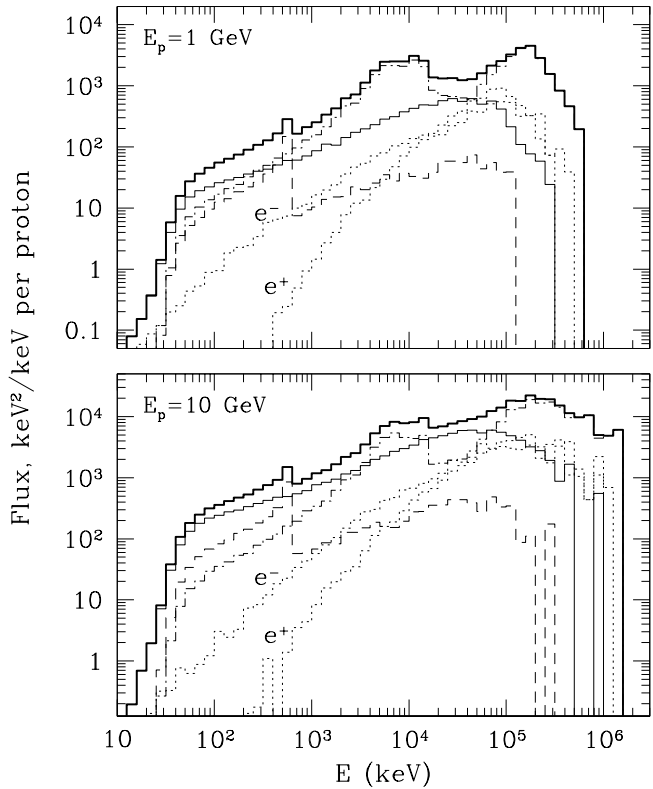} 
\caption{{\sl Top:} Spectral energy distribution (in $E^2
dN_\gamma/dE$ units) of photons emergent from the atmosphere exposed
from the top to an isotropic distribution of protons of energy
$\ep=1$~GeV. The thick solid line shows the total spectrum while the
other lines show its different components according to the photon
production process: free-free emission (solid), electron--positron
annihilation (dashed), other processes including nuclear deexcitation
and $\pi^0$ decay (dash-dotted). Also shown are the spectra ($E_{\rm
e}^2 dN_{\rm e}/dE_{\rm e}$) of secondary electrons and positrons
escaping from the atmosphere (dotted). {\sl Bottom:} The
same but for $\ep=10$~GeV. 
} 
\label{proton_components}
\end{figure}

\begin{figure}
\centering
\includegraphics[height=0.5\textheight]{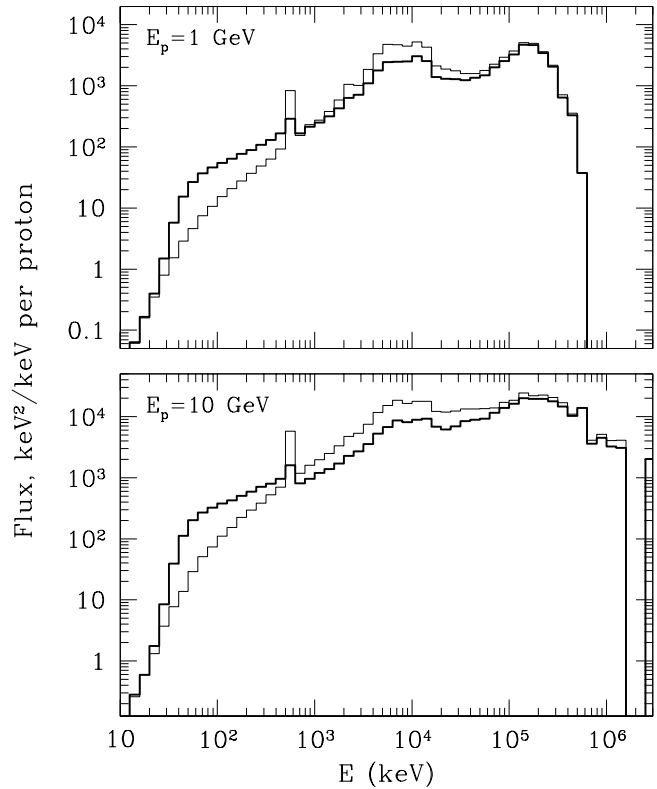}
\caption{{\sl Top:} Spectral energy distribution (in $E^2
dN_\gamma/dE$ units, thick solid line) of photons emergent from the
atmosphere exposed from the top to an isotropic distribution of
protons of energy $\ep=1$~GeV. The thin line shows the spectrum of the
initial energies (i.e. before Comptonization) of the escaping
photons. {\sl Bottom:} The same but for $\ep=10$~GeV. 
} 
\label{proton_origin}
\end{figure}

Figures~\ref{proton_components} and \ref{proton_origin} show the
emergent zenith angle-integrated photon spectra for 
$\ep=1$ and 10~GeV. One can see that several photon production
mechanisms provide comparable contributions to the outgoing 
spectra. These include: i) free-free emission by electrons and
positrons (distributed over several orders of magnitude in photon energy), ii)
electron--positron annihilation (mostly in a line at 511~keV) and iii)
high-energy processes including nuclear deexcitation (in the $\sim$~1--15~MeV 
range\footnote{Note that because of the short lifetimes
(typically a tiny fraction of a second) of the excited state,
atmospheric gamma-ray lines may be considered prompt (see  
e.g. \citealt{rametal79}.)} and $\pi^0$ decay (a broad bump near 
100~MeV). The energy albedo of the atmosphere (ratio of the emergent
radiation flux over the incident energy flux of cosmic rays) ranges
from $\sim 1$\% for 1~GeV protons to $\sim 0.1$\% for 250~GeV protons.  

As is evident from Fig.~\ref{proton_origin}, most of the
escaping photons with final energies below $\sim 511$~keV have
experienced single or multiple Compton down-scatterings and thus lost
a significant or dominant fraction of 
their initial energy due to electron recoils. The emergent hard X-ray spectrum
has thus lost much of the memory about its parent emission processes 
and as a result is fairly similar to the Comptonized spectra appearing
in various astrophysical situations where hard X-rays or gamma-rays
propagate through a dense and cold environment, e.g. in supernovae
\citep{gresun87}. In particular, such a spectrum was observed from SN
1987A \citep{sunetal87}.

\subsubsection{Adequacy of the adopted atmosphere model}
\label{s:simul_proton_setup}

\begin{figure}
\centering
\includegraphics[width=\columnwidth]{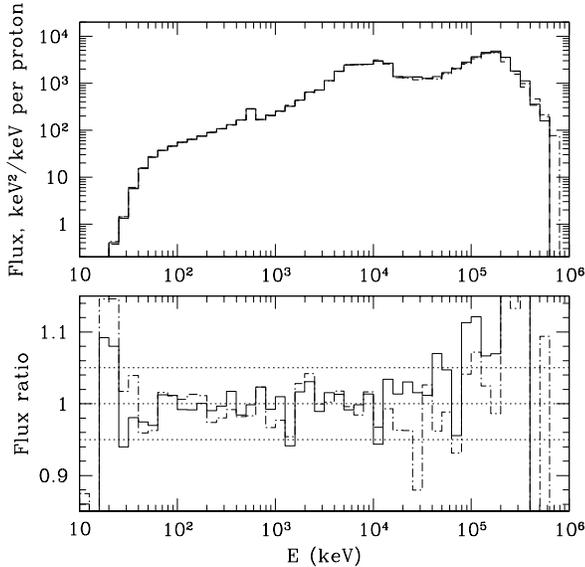}
\caption{{\sl Top:} Spectral energy distribution (in $E^2
dN_\gamma/dE$ units) of photons emergent from the
atmosphere exposed from the top to an isotropic distribution of
protons of energy $\ep=1$~GeV for different models of the atmosphere:
a slab geometry with a total height of 96~km (the reference model) --
the dashed line, a spherical shell geometry with an inner radius of
6400~km and the same vertical structure as the reference model -- the
dashed line, and a slab geometry with a total height of 48~km and the
same vertical structure as the lower half of the reference model -- the
dash-dotted line. {\sl Bottom:} The ratio of the spectra obtained for 
the spherical and 48-km slab models to that obatined for the reference
model -- the solid and dash-dotted line, respectively.
} 
\label{proton_setups}
\end{figure}

Our using of a plane-parallel model for the Earth's atmosphere is
motivated by the fact that all of typical free paths of particles and photons
in the problem under consideration are less than a few ten~km and are
thus much shorter than the Earth radius. Nevertheless, to verify the
adequacy of the adopted approximation we performed a series of more realistic
simulations, in which the Earth's atmosphere was modeled as a
spherical shell with an inner radius of 6400~km and an outer radius of
6496~km. The vertical structure and chemical composition of the
atmosphere were adopted exactly the same as in our reference (slab-geometry)
model. In Fig.~\ref{proton_setups} we compare the angular-integrated
emergent spectrum produced by 1~GeV protons in the spherical
atmosphere with that produced in the plane-parallel one. As expected,
there is no significant difference between these two spectra apart
from the statistical noise. We found this to be the case also for
other incident proton energies. These results confirm that the
plane-parallel approximation is appropriate for our study (see also
Sect.~\ref{s:simul_proton_angle} below).

We also show in Fig.~\ref{proton_setups} a simulated spectrum for a
plane-parallel atmosphere that has the same sea-level density
(1.2~mg~cm$^{-3}$) and scale-height (8~km) as our reference model but
is only 48~km (instead of 96~km) high. In this case too there are no
significant deviations in the outgoing spectrum relative to the one computed 
within our standard model. We present this example to demonstrate the
very weak sensitivity of the emergent spectrum to changes in the setup
of the atmosphere other than its chemical composition, which was
mentioned at the beginning of Sect.~\ref{s:simul}.

In addition, Fig.~\ref{proton_setups} provides a sense of the
statistical uncertainties pertaining to our simulations, which, as we
have already mentioned, are at the level of a few per cent over the
25~keV--10~MeV energy range.

\subsubsection{Fitting formulae for the emergent hard X-ray spectrum}
\label{s:simul_proton_fits}

\begin{figure}
\centering
\includegraphics[width=\columnwidth]{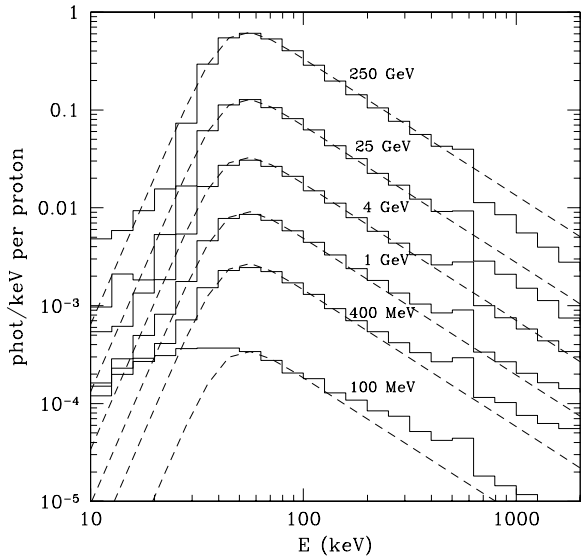}
\caption{Emergent spectra (solid lines, note the different
units compared to the previous figures) in the X-ray to soft gamma-ray 
range for different energies of incoming cosmic protons (labels next
to the curves). The dashed lines show the analytic approximation for the
25--300~keV range given by equation~(\ref{eq:spec_fit}). This
approximation breaks down at $\ep\lesssim$400~MeV. 
} 
\label{proton_spectra}
\end{figure}

On performing many simulations, we proved (see
Fig.~\ref{proton_spectra}) that the emergent spectrum between 25 and
300~keV is barely sensitive to the energy 
of the parent cosmic proton in the range of interest to us, namely from
$\sim 400$~MeV to $\sim 250$~GeV (see Sect.~\ref{s:predict}). The outgoing
photon spectrum over 25--300~keV can be well approximated by the following
fitting formula:

\begin{equation}
\frac{dN_{\gamma}}{dE}=\frac{C}{(E/44~{\rm keV})^{-5}
+(E/44~{\rm keV})^{1.4}}\,\,\,{\rm keV}^{-1},
\label{eq:spec_fit}
\end{equation}
where the coefficient $C$ depends on the primary particle energy as is
discussed below.

The photon spectrum (Fig.~\ref{proton_spectra}) peaks at
$\sim 50$--60~keV. At energies below 40~keV the spectrum exhibits
a rapid decline ($\propto E^5$) due to Compton down-scatterings. Above
$\sim 300$--400~keV, where the escaping photons 
have experienced at most few Compton scatterings, the spectrum
(including the strength of the 511~keV line) starts to exhibit a
noticeable dependence on the incident proton energy. This is also the
case for photons with energies less than $\sim 25$~keV, most of which
were born via bremsstrahlung in the skin 
of the atmosphere with surface density $\sim$ several g~cm$^{-2}$ and
whose spectrum retains signatures of the energy distribution of the
parent secondary electrons. We note that at these low energies
the emergent X-ray flux is very weak due to photoabsorption.

\begin{figure}
\centering
\includegraphics[width=\columnwidth]{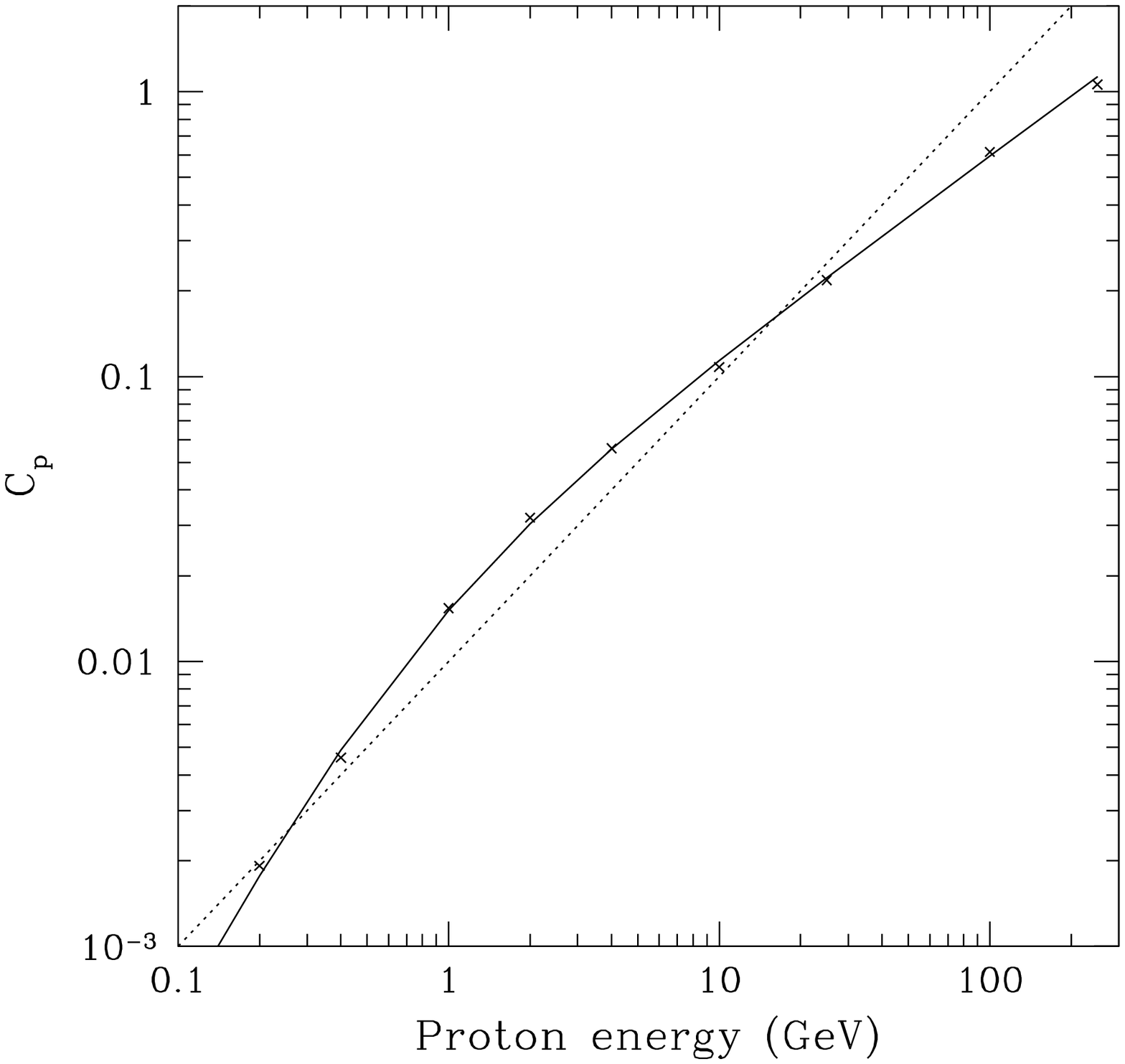}
\caption{Normalization of the emergent 25--300~keV spectrum,
equation~(\ref{eq:spec_fit}), as a function of proton energy
$\ep$. Shown are values derived from simulations (points) and the fitting curve
(solid) given by equation~(\ref{eq:proton_norm}). An arbitrarily
scaled linear dependence is shown for comparison (dotted).
} 
\label{proton_norm_energy}
\end{figure}

We next address the amplitude of the resulting photon
spectra. In Fig.~\ref{proton_norm_energy} the coefficient $C$
appearing in equation~(\ref{eq:spec_fit}) is shown as a function of the
incident proton energy $\ep$. The results of simulations in the range
400~MeV--250~GeV are well described by the following expression:
\beq
C_{\rm p}(\ep)=\frac{0.0166}{(\ep/0.6~{\rm GeV})^{-1.8}+(\ep/0.6~{\rm
GeV})^{-0.7}},
\label{eq:proton_norm}
\eeq
which allows one to calculate the emergent intensity for a given
incident spectrum of cosmic protons, as we do in \S\ref{s:predict} below.

\subsubsection{Angular dependence of emergent radiation}
\label{s:simul_proton_angle}

\begin{figure}
\centering
\includegraphics[width=\columnwidth]{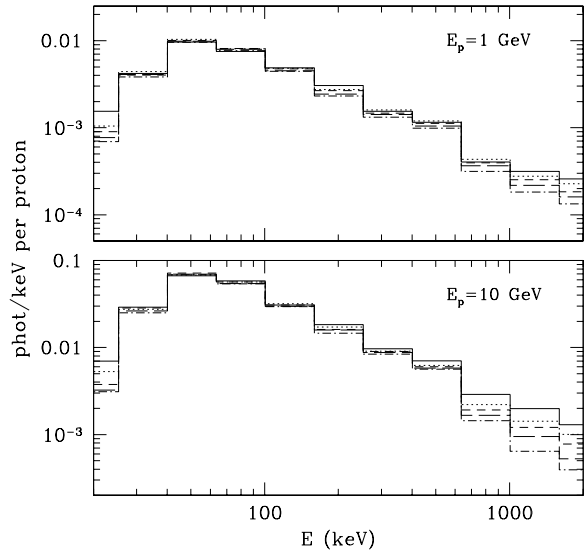}
\caption{{\sl Top:} Spectra for $\ep=1$~GeV, emergent at
different zenith angles ($\mu=\cos\theta$):
$\mu=$0--0.2 (solid), 0.2--0.4 (dotted), 0.4--0.6 (short-dashed),
0.6--0.8 (long-dashed) and 0.8--1.0 (dash-dotted). The spectra were divided by
$\mu(1+\mu)$. {\sl Bottom:} The same but for $\ep=10$~GeV.
} 
\label{proton_directions}
\end{figure}

\begin{figure}
\centering
\includegraphics[width=\columnwidth]{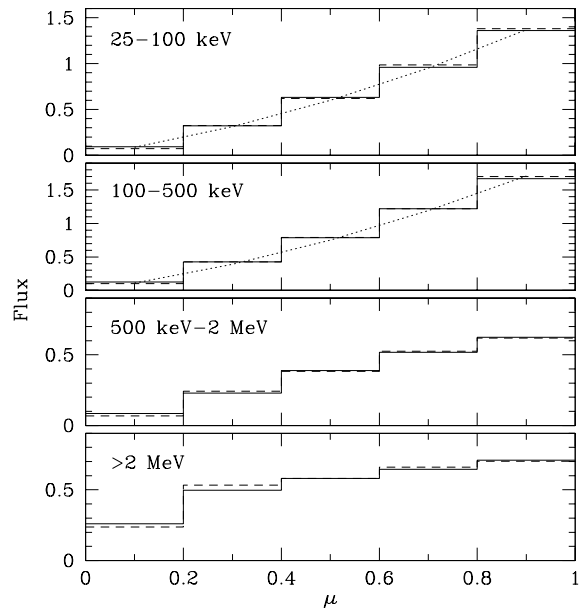}
\caption{Emergent flux for $\ep=10$~GeV as a function of the zenith
angle ($\mu=\cos\theta$) in different photon energy ranges:
25--100~keV, 100--500~keV, 500~keV--2~MeV and above 2~MeV. The solid
and dashed lines show the results obtained using the plane-parallel
and spherical atmosphere models. In the upper two
panels also a well-matching $\sim\mu(1+\mu)$ dependence is shown (dotted).
} 
\label{flux_directions}
\end{figure}

We proceed to compare spectra emergent at different zenith angles. We
found (see Fig.~\ref{proton_directions}) that the 
spectral shape between 25 and 300~keV remains stable and
as given by equation~(\ref{eq:spec_fit}), while the flux
(Fig.~\ref{flux_directions}) approximately obeys the law
\beq
dF\propto\mu(1+\mu)d\mu.
\label{eq:flux_angular}
\eeq

This means that the outgoing hard X-ray radiation exhibits some central
brightening as compared to the classical Lambert case, $dF\propto\mu
d\mu$, corresponding to black-body radiation. The angular distribution
of the atmospheric hard X-ray emission is also different from the
law of darkening pertaining to the other classical case of a
semi-infinite scattering atmosphere with radiation coming from deep
interiors: $dF\propto\mu(1+2\mu)d\mu$ \citep{chandra,sobolev}. The
resulting distribution actually resembles that of the Comptonized
radiation escaping from a purely scattering disk with optical thickness
$\tau_{\rm T}\sim$~a few \citep{suntit85}. This reflects the fact that
although cosmic rays generate radiation over several hundred
g~cm$^{-2}$ deep into the atmosphere, only radiation produced in the top
few ten g~cm$^{-2}$ of air can efficiently survive multiple Compton
down-scatterings (as well as photoabsorption) and escape into space in
the form of hard X-rays.

On the contrary, the emergent gamma-rays (with energies
$\gtrsim$~1--2~MeV) exhibit some limb brightening (enhanced flux at
small $\mu$'s and reduced flux at large $\mu$'s), see the bottom panel
of Fig.~\ref{flux_directions}), which can be attributed to the fact
that escaping gamma quanta have experienced at most a few Compton scatterings.

In Fig.~\ref{flux_directions} we also compare the result obtained
using our standard atmosphere model with one computed using the
spherical setup of the atmosphere, described in
Sect.~\ref{s:simul_proton_setup}. Similarly to the angular-integrated 
spectrum, there are no significant differences between the two derived
angular dependences of the emergent flux. Specifically, the 
difference between the two approximations is less than 3\% of the average
flux, $\langle dF/d\mu \rangle$, in each bin including the $0<\mu<0.2$
one (corresponding to zenith angles between 78.5$^\circ$ and
90$^\circ$). We do expect though the plane-parallel 
approximation to eventually break down at grazing angles approaching
90$^\circ$. However, at these angles only a tiny fraction of the total flux
from the Earth is emitted (for example the fraction of the total flux emitted
at $\mu<0.2$ is $\sim 3$\%), which implies that our plane-parallel
model should be applicable to virtually any space-born experiment.

\subsection{Alpha particles}

\begin{figure}
\centering
\includegraphics[width=\columnwidth]{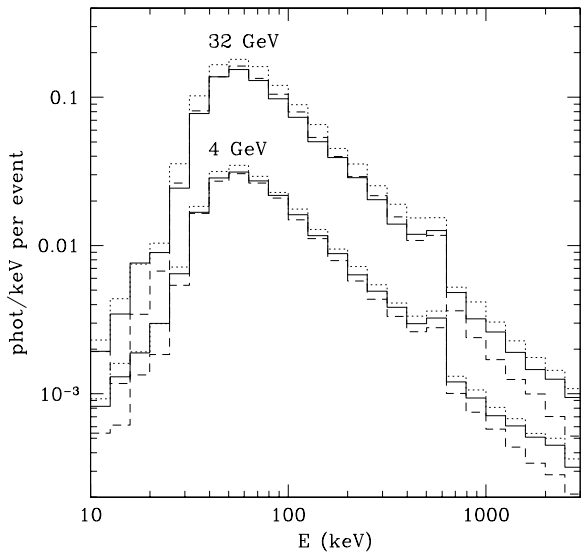}
\caption{Emergent spectra in the X-ray to soft gamma-ray 
range generated by $\alpha$-particles of energy $E_\alpha=4$ and
32~GeV (solid lines) compared to the spectra generated by
protons of energy $E_{\rm p}=E_\alpha$ (dashed lines) and by four
protons each of energy $E_{\rm p}=E_\alpha/4$ (dotted lines).
} 
\label{alpha_proton}
\end{figure}

A major component of Galactic cosmic rays is He nuclei. {\sl
GEANT4} permits to handle inelastic scatterings of $\alpha$-particles
up to $E_\alpha\sim 40$~GeV. We therefore ran a number of simulations of
bombardment of the atmosphere by monoenergetic
$\alpha$-particles and found (see Fig.~\ref{alpha_proton}) that the 
resulting hard X-ray spectra are nearly identical to those induced by
protons of the same energy as the $\alpha$-particles ($E_{\rm
p}=E_\alpha$). If the spectrum generated by an $\alpha$-particle of
energy $E_\alpha$ is instead compared with that generated by four
protons each of energy $E_{\rm p}=E_\alpha/4$ (a possibility suggested
by the simplest superposition theory), then the match becomes somewhat
worse. We can therefore extend the use of equation~(\ref{eq:proton_norm}),
originally derived for protons, to $\alpha$-particles, in which
case one should simply substitute $E_\alpha$ for $E_{\rm p}$ in the formula. 

\subsection{Cosmic electrons and positrons}

\begin{figure}
\centering
\includegraphics[width=\columnwidth]{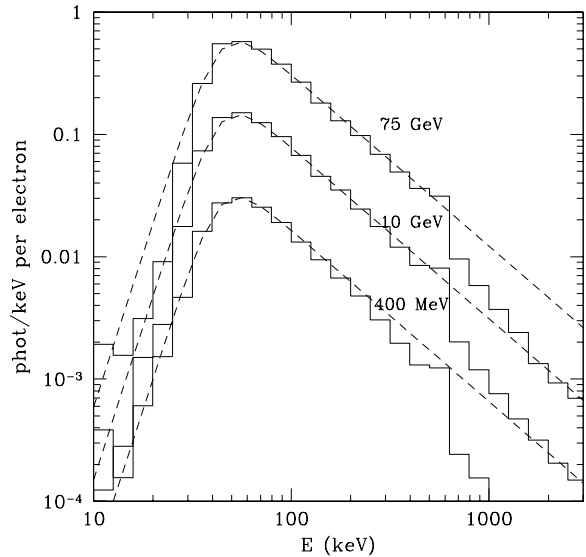}
\caption{Emergent spectra (solid lines) in the X-ray to soft gamma-ray
range generated by cosmic electrons of different energies. The dashed
lines show the analytic approximation in the 25--300~keV range given by
equation~(\ref{eq:spec_fit}).  
} 
\label{electron_spectra}
\end{figure}

A significant fraction of Galactic cosmic rays consists of
electrons (with a small fraction of positrons). We therefore also
performed simulations for monoenergetic electrons. The resulting hard X-ray
spectra are not significantly different from the case of protons and
in the energy band 25--300~keV equation~(\ref{eq:spec_fit}) remains a good
approximation of the spectra (Fig.~\ref{electron_spectra}). We
also tested positrons and, as expected, obtained nearly identical results.

\begin{figure}
\centering
\includegraphics[width=\columnwidth]{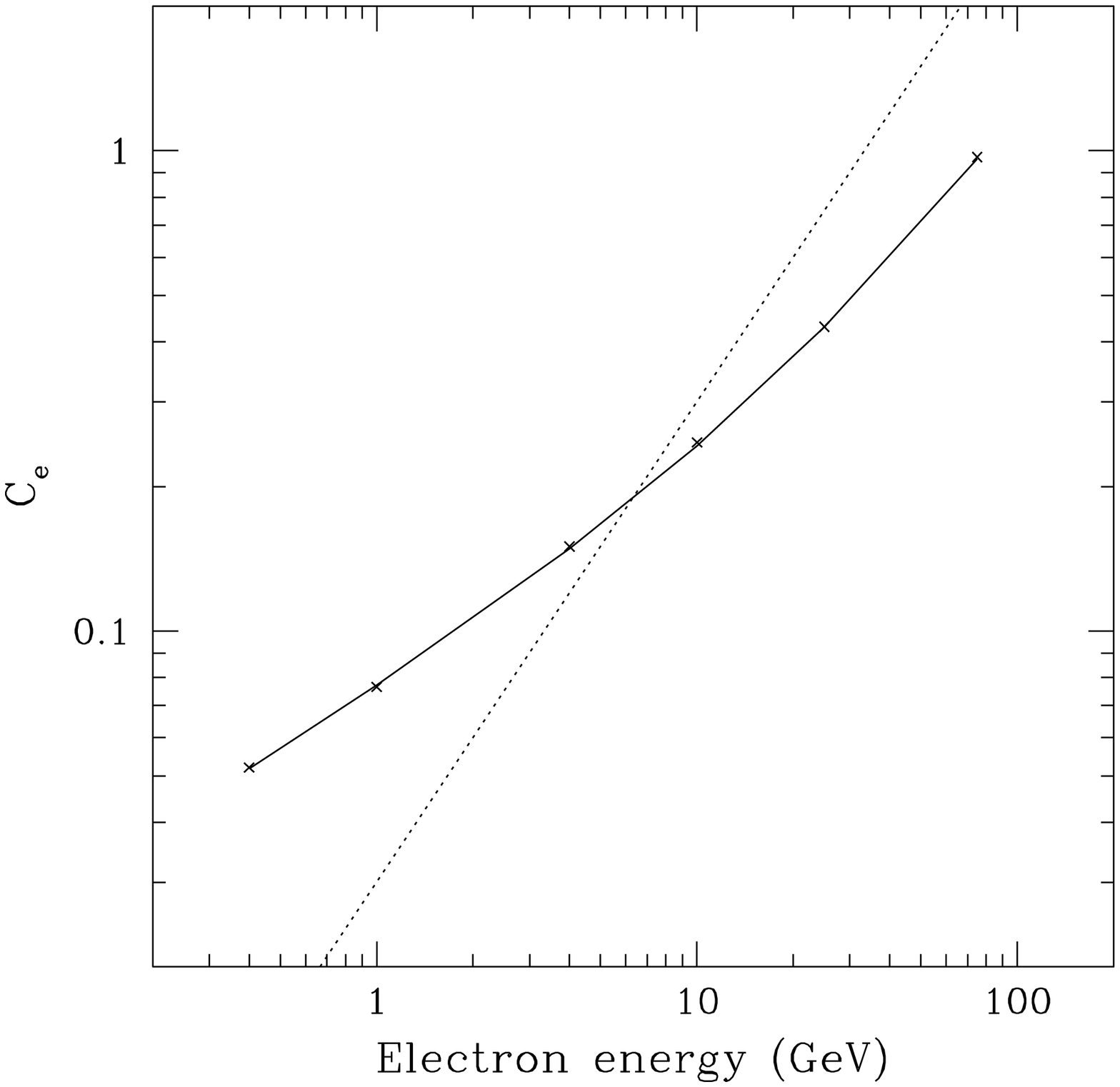}
\caption{Normalization of the emergent 25--300~keV spectrum,
equation~(\ref{eq:spec_fit}), as a function of electron energy
$\ee$. Shown are values derived from simulations (points) and the fitting
curve (solid) given by equation~(\ref{eq:electron_norm}). An
arbitrarily scaled linear dependence is shown for comparison (dotted).
} 
\label{electron_norm_energy}
\end{figure}

In Fig.~\ref{electron_norm_energy} the spectral normalization $C_{\rm
e}$ -- to be substituted for $C$ in equation~(\ref{eq:spec_fit}) -- is shown
as a function of the electron energy $\ee$. This 
dependence can be approximated as
\beq
C_{\rm e} (\ee)=0.072\left(\frac{\ee}{\rm GeV}\right)^{0.4}
\left[1+0.07\left(\frac{\ee}{\rm GeV}\right)^{0.69}\right].
\label{eq:electron_norm}
\eeq

Similarly to the case of protons, in \S\ref{s:predict} below we integrate
equation~(\ref{eq:electron_norm}) over the
energy spectrum of cosmic electrons bombarding the Earth. 
  
\subsection{Re-entrant albedo particles}
\label{s:simul_reentr}

So far we have implicitly assumed that only energetic particles
coming from space can induce X-ray and gamma-ray emission in the
atmosphere. However, already early studies (e.g. \citealt{puskin70})
indicated that additional radiation should be generated by re-entrant albedo
particles, i.e. secondary electrons and positrons that were born in
air showers, escaped from the atmosphere but were then deflected
back into it by the geomagnetic field. Although an accurate evaluation of
this effect would require more detailed simulations involving magnetic
fields, it is possible to make a first-order estimate. For this
purpose, we used the recorded spectra of secondary electrons and
positrons induced by cosmic protons and emergent from the atmosphere
(Fig.~\ref{proton_components}) as new input spectra for bombarding the
atmosphere. This is a reasonable procedure since the low magnetic
rigidities (momenta per charge) of most of these particles ($< 1$~GV)
do not allow them to escape into interplanetary space. 

\begin{figure}
\centering
\includegraphics[width=\columnwidth]{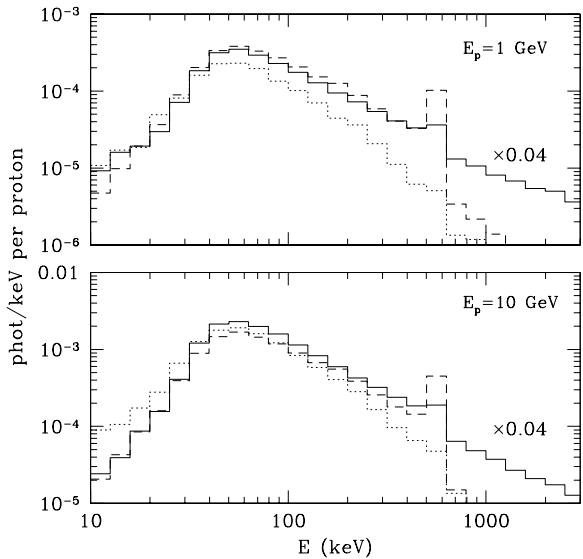} 
\caption{{\sl Top:} Emergent spectra produced by the
re-entrant albedo electrons (dotted) and positrons (dashed) generated
by protons of energy $\ep=1$~GeV. The spectrum produced during the
primary cascade, multiplied by 0.04, is shown for comparison (solid
line). {\sl Bottom:} The same but for $\ep=10$~GeV. 
} 
\label{reentr}
\end{figure}

In Fig.~\ref{reentr} we show the emergent photon spectra
generated by re-entrant particles in comparison with the previously
computed spectrum formed during the primary proton-induced
cascade. It can be seen that the spectral shape of the secondary
components produced by re-entrant electrons and positrons in the
25--300~keV range is not significantly different from that of the
primary spectrum and that re-entrant particles contribute together
$\sim 8$\% to the hard X-ray flux. Morever, we verified that this
amount depends very little on the primary proton energy. We take
this minor additional component into account when making our final
predictions in \S\ref{s:predict}.

We also explored the contribution of re-entrant albedo particles generated by
primary cosmic electrons. The corresponding additonal hard X-ray
flux proves to be less than 1\% of the flux generated by the primary
electrons, i.e. $\lesssim 10^{-3}$ of the flux due to cosmic protons,
and thus can be neglected. 

\section{Predictions for observations}
\label{s:predict}

In the preceeding section we demonstrated that cosmic
particles of different types and energies generate in the Earth's
atmosphere hard X-ray radiation with a nearly universal spectrum,
which can be approximately described by equation~(\ref{eq:spec_fit}). We also 
determined the dependence of the emergent flux on the primary particle
energy. On the other hand, the spectrum and chemical composition of
Galactic cosmic rays impinging on the Earth\footnote{Note that after
solar flares the Earth is bombarded by additional cosmic rays with
variable spectrum. This transient component is not considered here.}
are known to a good accuracy, in particular thanks to the many
experiments of the past decade. The spectrum is stable at high
energies (above $\sim 10$~GeV per nucleon) but varies significantly at
lower energies over the 11-year solar cycle due to modulation by the
interplanetary magnetic field and solar wind, as well as with position
on the Earth due to a rigidity cutoff imposed by the terrestrial
magnetosphere. Since these effects are fairly well understood, we can
combine the available information about the cosmic-ray 
spectrum with the results of our simulations for monoenergetic
particles (\S\ref{s:simul}) to make predictions for observations of
the Earth's hard X-ray radiation by satellite-born instruments.

\subsection{Integrating over the cosmic-ray spectrum}

To describe the spectrum of cosmic protons as well
as its dependence on the solar modulation potential $\phi$ ($Ze\phi$
corresponds to the average energy loss of cosmic rays inside the
heliosphere, where $Z$ is the charge number of a particle and $e$ is
the electron charge), we use the approximation provided by
\citet{usoetal05}, which is based on a summary of recent
measurements. Specifically, we adopt for the differential spectrum of
protons at 1~AU 
\beqa
J_{\rm p}(E,\Phi) &=&
\frac{1.9[(E+\Phi)
(E+\Phi+1.876)]^{-1.39}}
{1+0.4866[(E+\Phi)(E+\Phi+1.876)]^{-1.255}}
\nonumber\\
&&\times
\frac{E(E+1.876)}
{(E+\Phi)(E+\Phi+1.876)}
\nonumber\\
&&{\rm particles}
~({\rm s~cm}^2~{\rm sr~GeV/nucleon})^{-1},
\label{eq:protons}
\eeqa
where $E$~(GeV) is the kinetic energy per nucleon and
$\Phi\,({\rm GeV})=(Ze/A)\phi$ (here $A$ is the atomic mass), so that
for protons $E=E_{\rm p}$ and 
$\Phi=e\phi$. We refer the reader to the above-mentioned paper (see 
also e.g. \citealt{gaietal01}) for references to the original data from the
experiments AMS, BESS, CAPRICE, IMAX, MASS, etc, most of which agree with
each other within $\sim 10$\% at energies below $\sim 200$~GeV. 

The Galactic cosmic-ray spectrum is additionally modified by the
geomagnetic field. To a first approximation, adopted 
below, this modulation can be represented by a sharp cutoff below a
certain value $\rcut$ (GV) of the magnetic rigidity of a charged 
particle, whose dependence on the geomagnetic coordinates is well-known. In
\S\ref{s:beams} below we estimate the systematic error associated
with this approximation.

Proceeding with our analysis, we integrated the approximation for
$C_{\rm p}$ given by equation (\ref{eq:proton_norm}) over the spectrum
of cosmic protons given by equation~(\ref{eq:protons}) for specified
values of the solar modulation potential $\phi$ and geomagnetic cutoff
rigidity $\rcut$. The integration was 
done up to $\ep=1000$~GeV, protons with energies higher than 250~GeV
contributing $\sim 3$\% to the emergent radiation flux for the highest
expected values of $\rcut\sim 15$~GV and less for lower cutoff values. 

\begin{figure}
\centering
\includegraphics[height=0.5\textheight]{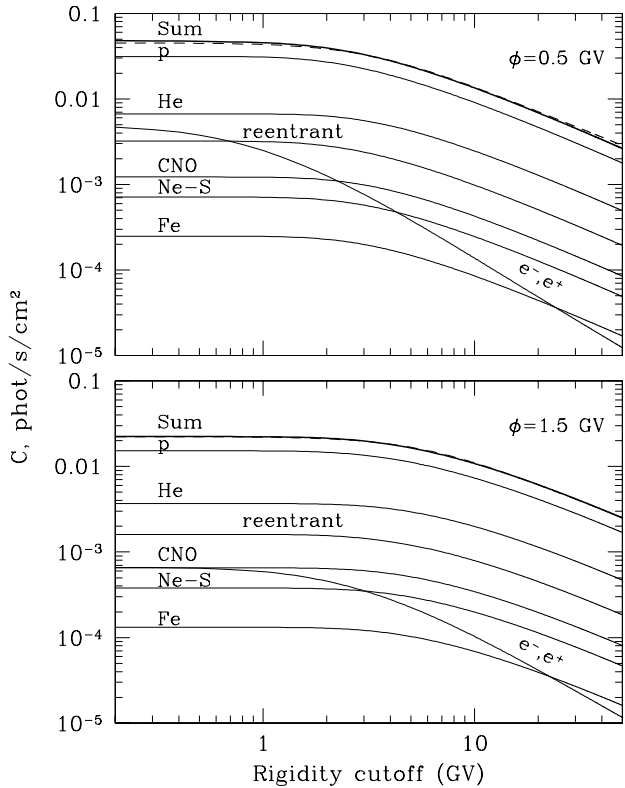}
\caption{{\sl Top:} Predicted hard X-ray surface brightness
(integrated over all zenith angles) of the Earth's
atmosphere during solar minimum ($\phi=500$~MV) as a function of the
geomagnetic cutoff rigidity (for a proton) $\rcut$. The thick solid
line shows the expected total signal -- coefficient $C$ in the
spectral approximation given by equation~(\ref{eq:spec_fit}), which
consists of several components (thin solid lines): due to protons,
$\alpha$-particles, CNO nuclei, Ne-S nuclei, Fe nuclei, cosmic
electrons and positrons, and re-entrant electrons and positrons. The
cumulative flux is well fit by equation~(\ref{eq:norm_fit}) -- the
dashed line nearly coincident with the thick solid line. {\sl Bottom:}
The same but for solar maximum ($\phi=1.5$~GV). 
} 
\label{norm_rigcut}
\end{figure}

In Fig.~\ref{norm_rigcut} we show the predicted hard X-ray 
flux induced by cosmic protons as a function of $\rcut$ for two values
of $\phi$: 0.5 and 1.5~GV, which are typical for solar minima
and maxima, respectively (when the suppression of the low energy
cosmic-ray flux is weakest and strongest, respectively). One can see that the
emergent hard X-ray flux is almost independent of $\rcut$ when
$\rcut\lesssim 1$~GV, as could be expected since the flux of incident
protons with rigidities below $\sim 1$~GV is strongly suppressed by
solar modulation. As follows from Fig.~\ref{norm_rigcut}, the emergent
hard X-ray flux is expected to increase by a factor of $\sim 2$ during
solar minima compared to solar maxima in regions with low
geomagnetic cutoff rigidity, i.e. near the magnetic poles of the
Earth. The solar-cycle variations become progressively smaller with
increasing cutoff rigidity and reduce to $\sim 10$\% near the geomagnetic
equator, where $\rcut\sim 15$~GV.   

We next performed a similar calculation for $\alpha$-particles. In
this case the Galactic spectrum is very similar to that of protons,
while He composes $\sim 5$\% (in particle number) of cosmic rays. The
solar-modulated spectrum of $\alpha$-particles can thus be well
approximated by equation~(\ref{eq:protons}) multiplied by 0.05
\citep{usoetal05}. Note that since for a given particle energy the rigidity
of $\alpha$-particles is a factor of $\sim 2$ smaller than that of
protons, the effects of the solar and geomagnetic modulation are
accordingly modified. The results of \S\ref{s:simul} indicate that the
radiation flux produced by an $\alpha$-particle of given energy is
nearly equal to that induced by a proton of the same energy (at least
at energies below 40~GeV). This makes the corresponding calculation 
straightforward. In Fig.~\ref{norm_rigcut} we show as a function of
$\rcut$ (geomagnetic cutoff rigidity for a proton) the predicted hard
X-ray surface brightness of the Earth's atmosphere due to bombardment
by $\alpha$-particles during solar minimum and maximum. The hard X-ray
flux induced by $\alpha$-particles proves to be $\sim 22$\% and
$\sim 28$\% at $\rcut\lesssim 1$~GV and $\rcut\gg 1$~GV, respectively, of
that due to protons.

We can estimate in the same manner  the contribution of heavier cosmic
particles. To this end we assume that atoms of the groups CNO,
Ne-S and Fe have the same Galactic spectra as $\alpha$-particles and protons,
while their differential fluxes (per GeV/nucleon) are 6.8\%, 2.6\% 
and 0.54\% of that of $\alpha$-particles
(e.g. \citealt{fioetal01}). Since for these groups of atoms the
relation $Z=A/2$ approximately holds, the helio- and geomagnetic 
effects for them are similar to the case of He. Assuming that,
similarly to the case of He, the hard X-ray flux induced by a heavy
particle is the same as that produced by a proton of equal
energy, the hard X-ray surface brightness of the Earth's 
atmosphere due to bombardment by heavy cosmic particles can be readily
estimated (Fig.~\ref{norm_rigcut}).

We can next estimate the hard X-ray surface brightness of the
atmosphere due to bombardment by cosmic electrons and positrons. In
this case we should use equation (\ref{eq:electron_norm}) instead of
equation (\ref{eq:proton_norm}). Also the magnetic rigidity effects
are somewhat different from the case of protons and nuclei because the
rest energy of electrons and positrons is negligible. Adopting for the
solar-modulated spectrum of cosmic electrons \citep{webber83}
\beq
J_{\rm e}(E_{\rm e},\phi)=
0.07(E_{\rm e}+e\phi)^{-3.3}\left(\frac{E_{\rm e}}{E_{\rm e}+e\phi}\right)^2
\,\,\,({\rm s~cm}^2~{\rm sr~GeV})^{-1},
\label{eq:electrons}
\eeq
we obtain the dependence of the emergent
hard X-ray flux on $\rcut$ shown in Fig.~\ref{norm_rigcut}.
 
Finally, the contribution of re-entrant albedo electrons
and positrons can be roughly estimated by multiplying the hard X-ray
flux due to cosmic protons and nuclei by 0.08 (see \S\ref{s:simul_reentr}),
since albedo particles are expected to re-enter the
atmosphere in the opposite geomagnetic hemisphere in regions with
similar geomagnetic cutoff rigidity to their escape positions. This
emission component as well as the total atmospheric signal are
shown as functions of $\rcut$ in Fig.~\ref{norm_rigcut}.

\subsection{Approximate formula for the hard X-ray surface brightness of the
Earth's atmosphere} 

We found that over the region of practical interest: $0.3<\phi ({\rm
GV}) <2$ and $\rcut ({\rm GV})\lesssim 50$ -- the 
dependence of the atmosphere's hard X-ray surface brightness on the
solar modulation potential $\phi$, geomagnetic cutoff rigidity
$\rcut$ and zenith angle $\theta$ ($\mu=\cos\theta$) can be well
described by the following formula: 
\beqa
C=\frac{3\mu(1+\mu)}{5\pi}
\frac{1.47\times 0.0178\left[(\phi/2.8)^{0.4}+(\phi/2.8)^{1.5}\right]^{-1}}
{\left\{1+\left[\rcut/(1.3\phi^{0.25}(1+2.5\phi^{0.4}))\right]^2\right\}^{0.5}}
\nonumber\\
({\rm s~cm}^2~{\rm sr})^{-1}.
\label{eq:norm_fit}
\eeqa
To calculate the emergent hard X-ray spectrum (from 1~cm$^{2}$ of the
Earth's atmosphere into a unit solid angle) for given values of
$\phi$, $\rcut$ and $\mu$, one just needs to substitute the amplitude $C$
given by equation~(\ref{eq:norm_fit}) into
equation~(\ref{eq:spec_fit}).

We first derived the above dependence for protons only
and then proved that the combined contribution of the other cosmic-ray
constituents, despite their somewhat different dependences on $\rcut$ (see
Fig.~\ref{norm_rigcut}), can be well represented by the additional factor 1.47
appearing in equation~(\ref{eq:norm_fit}). This factor also
takes into account the estimated $\sim 8$\% contribution of re-entrant
albedo particles.

\begin{figure}
\centering
\includegraphics[width=\columnwidth]{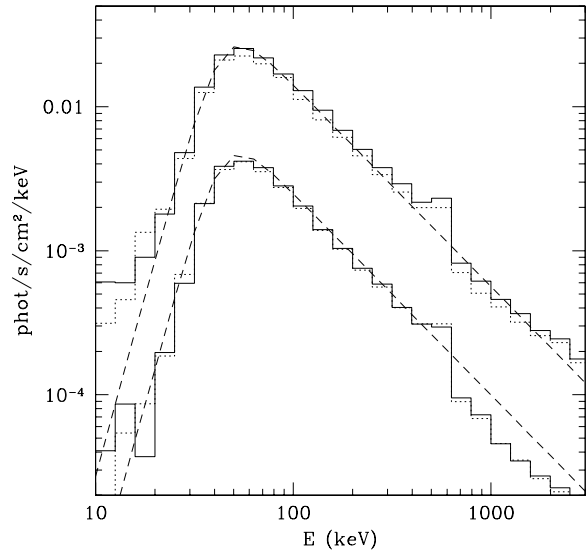}
\caption{Simulated spectra (solid lines) of Earth atmospheric emission 
induced by cosmic protons for $\phi=0.5$~GV, $\rcut=0.5$~GV and
$\phi=1.5$~GV, $\rcut=15$~GV (upper and lower curves,
respectively). The dashed lines show the corresponding predictions for
the 25--300~keV energy range based on equations~(\ref{eq:spec_fit})
and (\ref{eq:norm_fit}) (integrated over all zenith angles). The
dotted lines are simulations based on the incident direction-dependent cutoff 
rigidity approach in the dipole approximation for the geomagnetic
field ($\rcut=0.5$~GV and $\rcut=15$~GV approximately correspond to
$\lambda_{\rm m}=64.6^\circ$ and $0^\circ$, respectively). All the
simulated spectra were multiplied by 1.47 to account for the
non-proton cosmic rays and re-entrant albedo particles.
} 
\label{cosmic_protons}
\end{figure}

To verify our prediction for the atmospheric hard X-ray spectrum given
by equations~(\ref{eq:spec_fit}) and (\ref{eq:norm_fit}), we 
performed a series of simulations where the atmosphere was bombarded
by the cosmic proton spectrum expected for specified ($\phi$, $\rcut$) values
according to equation~(\ref{eq:protons}). In
Fig.~\ref{cosmic_protons} we present two extreme examples of the
resulting emission spectra -- one corresponding to observing a
geomagnetic pole during solar minimum and the other to observing the
geomagnetic equator during solar maximum, in both cases on integrating
the outgoing radiation over all zenith angles. It can be seen
that the computed spectra are in very good agreement with our analytic 
prediction in the 25--300~keV energy range. This confirmes the
self-consistency of our treatment and the stated few per cent
statistical accuracy of our simulations.

\subsection{Effect of the cutoff rigidity's zenith angle dependence}
\label{s:beams}

In the preceeding discusssion we assumed that at a given position
on the Earth the geomagnetic cutoff rigidity is the same for all
directions of incidence. In reality, the effects of 
the geomagnetic field can be better represented by a cutoff rigidity
that also depends on the zenith ($\theta$) and azimuthal ($\xi$)
angles. If the geomagnetic field is approximated by a dipole, then the
cutoff rigidity is given by Stormer's formula (see e.g. \citealt{lipetal98}):
\beq
\rcut=\frac{M_\oplus}{2r^2}\frac{\cos^4\lambda_{\rm m}}
{[1+(1+\cos^3\lambda_{\rm m}\sin\theta\sin\xi)^{1/2}]^2},
\label{eq:stormer}
\eeq
where $M_\oplus\approx 8.1\times 10^{25}$~G~cm$^3$ is the magnetic
dipole moment, $r$ is the distance from the dipole center and
$\lambda_{\rm m}$ is the magnetic latitude. In the atmosphere,
i.e. near the Earth surface ($r\approx r_\oplus$),
$M_\oplus/(2r^2)\approx 59.4$~GV, so at the geomagnetic equator
($\lambda_{\rm m}=0$) $\rcut$ takes values between 10.2 and 59.4~GV,
the vertical ($\theta=0$) cutoff being 14.9~GV; at the poles $\rcut=0$
for all directions.

We ran a series of simulations where for given values of the solar
modulation potential $\phi$ and geomagnetic latitude $\lambda_{\rm m}$
the incident cosmic-proton spectra were dependent on the zenith and
azimuthal angles, namely they were cut off at $\rcut$ according to  
equation~(\ref{eq:stormer}). We then compared the produced spectra of
atmospheric hard X-ray emission with those
simulated assuming that regardless of the incident direction the
proton spectrum is truncated at the vertical cutoff value:
$\rcut(\lambda_{\rm m},\theta,\phi)=\rcut (\lambda_{\rm
m},\theta=0)$. Both kinds of spectra turned out to be mutually
consistent within a few per cent (see Fig.~\ref{cosmic_protons}), even
at the geomagnetic equator. We 
note however that Stormer's formula generally underestimates the
cutoffs because it allows penetration of charged particles with
trajectories that would have intersected the surface of the Earth
\citep{smashe04}. The emergent radiation flux may thus be smaller
than predicted here by a few per cent. More accurate simulations
should be based on a particle backtracking approach and use more realistic
models of the geomagnetic field.

\subsection{Observations of the Earth from space by a wide-field instrument}

\begin{figure}
\centering
\includegraphics[width=\columnwidth]{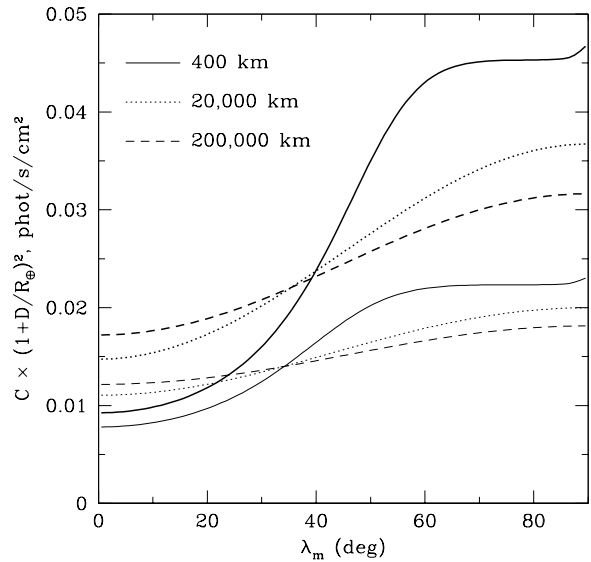}
\caption{Hard X-ray spectral flux -- coefficient $C$ in
equation~(\ref{eq:spec_fit}) -- that would be detected from the
whole Earth by a wide-field instrument in different circular orbits
(see notation on the figure), as a function 
of the spacecraft geomagnetic latitude. The thick and thin lines correspond to
observations during solar minimum ($\phi=0.5$~GV) and solar maximum
($\phi=1.5$~GV), respectively. Note that the dependencies were
rescaled as indicated along the vertical axis.
} 
\label{spacecraft_flux}
\end{figure}

Equation~(\ref{eq:norm_fit}) allows one to predict the hard X-ray
surface brightness of the atmosphere or its part when observed from a
spacecraft, provided that its orbit is
higher than $\sim$~50--100~km. As a practical example, we may use this
formula to predict the hard X-ray flux that would be measured by a
satellite-born instrument whose field of view covers the entire
terrestrial disk, which was the case in the recent INTEGRAL
observations. In the 
dipole approximation for the Earth's magnetosphere, the spectral
coefficient $C$ in equation~(\ref{eq:spec_fit}) can be calculated as a
function of the spacecraft geomagnetic latitude $\lambda_{\rm m}$ and
altitude $D$ (Fig.~\ref{spacecraft_flux}). One can see that in the
case of a low orbit ($D=400$~km in our example), during solar minimum the flux
detected during the passage of one of the magnetic poles
($\lambda_{\rm m}$) will be $\sim 5$~times higher than that measured 
during the passage of the equator. In the case of intermediate and
high orbits (20,000 and 200,000~km in our examples) the relative
amplitude of variations is expected to be $\sim 2$. 

We point out that since the geomagnetic field is not exactly dipolar, in
analyzing the data from a real experiment one should integrate
equation~(\ref{eq:norm_fit}) over the geomagnetic rigidity map
corresponding to a given observation. 

\section{Hard X-ray radiation from the Moon and planets}
\label{s:planets}

Obviously, the same mechanism of hard X-ray production should work in the
crust of the Moon, so it is interesting to consider this case as
well. Since the Moon has a negligible magnetic field, in its case both the 
incident cosmic-ray flux and the emergent hard X-ray flux
are expected to depend on the solar cycle phase only. 

\begin{figure}
\centering
\includegraphics[width=\columnwidth]{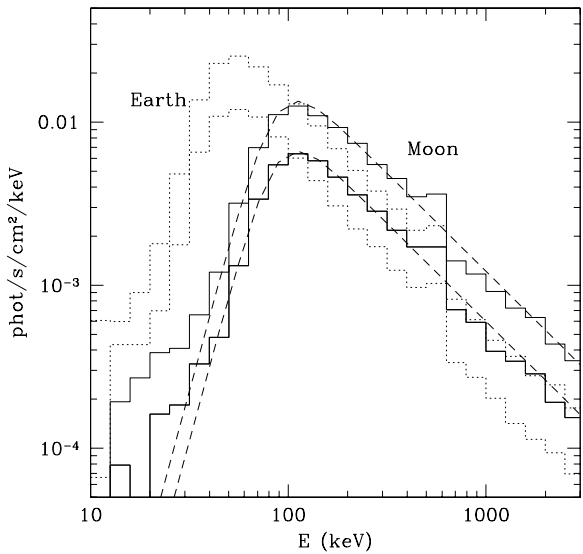}
\caption{Expected spectra of cosmic ray-induced emission from the
Moon's crust (integrated over all zenith angles) during 
solar minimum ($\phi=0.5$~GV) and maximum ($\phi=1.5$~GV) -- upper and
lower histogram, respectively. The dashed lines show the 
approximation in the 25--300~keV energy range given by
equation~(\ref{eq:fit_moon}). For comparison shown are the
corresponding spectra emitted near the magnetic poles of the Earth
($R_{\rm cut}=0$).  
} 
\label{planets}
\end{figure}

We performed simulations assuming that the Moon's crust has a density
of 3~g~cm$^{-3}$ and consists of O, Si, Al, Fe, Ca, Mg,
and Ti with mass fractions of 43, 21, 10, 10, 9, 5 and 2\%,
respectively. It turns out (see Fig.~\ref{planets}) that the spectrum emitted
above $\sim 100$~keV is similar to that produced in the Earth's
atmosphere and only the cutoff below 100~keV is stronger due to
photoabsorption on heavy atoms. The Moon's spectral surface brightness
can be approximated in the energy range 40--300~keV as follows: 
\beqa
\frac{dN_{\gamma}}{dE}
=\frac{3\mu(1+\mu)}{5\pi}\times 0.013
\left[\left(\frac{\phi}{2.8}\right)^{0.4}
+\left(\frac{\phi}{2.8}\right)^{1.5}\right]^{-1}
\nonumber\\
\left[\left(\frac{E}{88~{\rm keV}}\right)^{-4.5}
+\left(\frac{E}{88~{\rm keV}}\right)^{1.2}\right]^{-1}
\nonumber\\
({\rm s~cm}^2~{\rm sr}~{\rm keV})^{-1}.
\label{eq:fit_moon}
\eeqa
In writing down this formula we assumed, by analogy with the Earth's
atmosphere, that the total radiation flux produced in the Moon's crust
is that due to primary cosmic protons times 1.4. Note that the
corresponding factor for the Earth's atmosphere (1.47) included the
small contribution of re-entrant albedo particles.  

Similar results are expected also for Mars and Mercury, since the
chemical composition of their solid crusts is similar to that of the
Moon and they too have thin gaseous atmospheres and weak
magnetospheres. Also the effects of interplanetary magnetic fields on
the flux of cosmic rays are not expected to be significantly different
in the orbits of Earth, Mars and Mercury.
 
Interestingly, the Moon should appear as a fairly 
bright hard X-ray source when observed from a satellite orbiting the
Earth. Indeed, on integrating equation~(\ref{eq:fit_moon}) over the  
Moon disk one finds that a detector located at distance $D$
from the Moon will measure a spectral flux
\beqa
F_{\rm Moon}(D) &=&
(4\div 8)\times 10^{-7}\left(\frac{3\times 10^5~{\rm km}}{D}\right)^2 
\nonumber\\
&&\times
\left[\left(\frac{E}{88~{\rm keV}}\right)^{-4.5}
+\left(\frac{E}{88~{\rm keV}}\right)^{1.2}\right]^{-1}
\nonumber\\
&&\frac{{\rm photon}}{{\rm s~cm}^2~{\rm keV}},
\label{eq:moon_flux}
\eeqa
where the lower and upper values correspond to observations during
solar maximum and minimum, respectively. Therefore, a flux of $\sim
1$~mCrab is expected at 100--200~keV. 

One can expect hard X-ray production by cosmic rays to be much less
efficient on giant planets such as Jupiter and Saturn, since their typical 
cutoff rigidities (roughly proportional to the product of the magnetic
field strength on the surface of the planet and its radius) are much
higher than in the terrestrial case. Similarly, the spectrum of
Galactic cosmic rays impinging on the Sun is truncated at TeV
energies \citep{secetal91}, and this strongly suppresses the generation of
hard X-rays and gamma-rays in the solar atmosphere.

\section{Discussion and conclusions}

We have revisited the problem of the X-ray and gamma-ray emissivity of
the Earth's atmosphere taking advantage of the recent high-quality
measurements of the Galactic cosmic-ray spectra and the 
availability of efficient Monte Carlo codes incorporating
state-of-the-art descriptions of hadronic physics and electromagentic
cascades. We found that emission generated in the energy band
25--300~keV has a practically invariant spectral shape and a
flux that is expected to show significant variations both with
time (over the solar cycle) and with position on the Earth (due to
geomagnetic effects). We provided two simple expressions
-- equations~(\ref{eq:spec_fit}) and (\ref{eq:norm_fit}) -- which
allow one to predict the spectral intensity of atmospheric hard X-ray emission
for a particular observation from a spacecraft. It should be noted
that our results are not directly applicable to observations performed
from balloons floating at altitudes below $\sim$~50~km, since
then in addition to the upward flux there will also be a significant 
amount of downward radiation generated in the several overlying
g~cm$^{-2}$ of air.

The statistical accuracy of our predictions is better than 5\%, while the
systematic uncertainty is probably dominated by the current uncertainty
in our knowledge of the cosmic-ray flux, which is $\sim 10$\%. We can
try to assess the overall accuracy of our model through direct comparison with
real observations of the Earth from spacecraft. Among the few
published measurements of atmospheric emission in the hard X-ray
domain are those done in the 1970's from the satellites Kosmos~461
{\citep{goletal74} and 1972-076B \citep{imhetal76}. The spectra
obtained in the energy range $\sim$~50--1000~keV could be well
described by a power law with a photon index ($dN_\gamma/dE\propto
E^{-\Gamma}$) $\Gamma=1.55\pm 0.10$ in the former experiment and
$\Gamma\sim$~1.31--1.47 in the latter. These values are in good
agreement with our calculated value of 1.4 for the spectral slope
above $\sim 50$~keV. Furthermore, \cite{goletal74} (see also
\citealt{mazetal75}) measured the dependence of the hard X-ray flux on
geomagnetic cutoff rigidity over the range 
$3<\rcut~({\rm GV})<17.5$, and that result too appears to be
consistent with our computations. 

Recently during a scanning observation of the Earth, the INTEGRAL
observatory measured the spectrum of the cosmic X-ray background
together with that of atmospheric hard X-ray emission. Our analysis of the
data \citep{chuetal07} showed that the atmospheric spectrum at
energies from $\sim 40$ to $\sim 200$~keV is consistent with the
prediction of the present study, although the spectral shape was not well 
constrained by the data. When we adopted the spectral shape given
by equation~(\ref{eq:spec_fit}), we found the best-fit amplitude of
the atmospheric component to lie within 10\% of (actually only $\sim
4$\% higher than) the flux predicted by equation~(\ref{eq:norm_fit}) for the
relevant value of the solar modulation potential ($\phi=0.45$~GV) and
taking into account the geomagnetic cutoff ridigity map corresponding 
to the INTEGRAL observations \citep{chuetal07}.

This good agreement of our model with observations
suggests that the Earth could be used as an absolute
calibrator of future satellite-born sensitive hard X-ray
detectors, complementing the Crab nebula. This possibility as well as
the predictive power of our model could be tested soon via new INTEGRAL
observations of the Earth performed at a different phase of the solar cycle.

\begin{figure}
\centering
\includegraphics[width=\columnwidth]{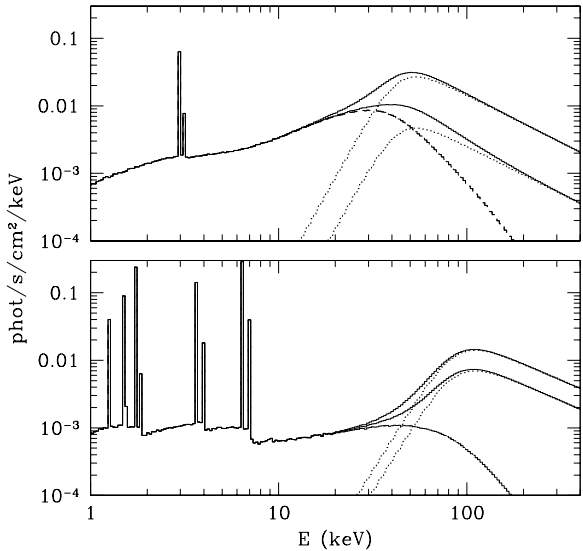}
\caption{{\sl Top:} Total X-ray spectral surface brightness of the
Earth (soild lines) -- sum of cosmic ray-induced
atmospheric emission (dotted lines) and the reflected CXB radiation
(dashed lines). The two sets of curves correspond to observations of a
geomagnetic polar region ($\rcut=0$) during solar minimum ($\phi=0.5$~GV)
and of the geomagnetic equator ($\rcut=15$~GV) during solar maximum
($\phi=1.5$~GV).  {\sl Bottom:} The same but for the Moon. The two sets of
curves correspond to observations performed during solar minimum and
solar maximum.  
} 
\label{planets_albedo}
\end{figure}

We also demonstrated that under the impact of cosmic rays the solid
surfaces of the Moon, Mars and Mercury should be about as
bright in hard X-rays as (in fact somewhat brighter than) the Earth's
atmosphere, making these objects potentially interesting for
observations. It is important to keep in mind though that in a real
experiment reflected CXB radiation will significantly contribute to
the observed signal at energies below $\sim 50$~keV. Based on the
computations of \citet{chuetal07b}, we show in
Fig.~\ref{planets_albedo} several examples of hard X-ray spectra that
could be observed from the Earth and Moon. This figure indicates that
the cosmic ray-induced hard X-ray continuum could complement the
CXB-induced fluorescent X-ray lines in probing the chemical
composition of the crusts of planets and the Moon, i.e. during the
Bepi Colombo mission to Mercury\footnote{http://sci.esa.int/home/bepicolombo}. 


\end{document}